\documentstyle[12pt,aaspp4]{article}

\def\ms{m\thinspace s$^{-1}$\ }

\begin{document}

\slugcomment{To appear in {\it Nature}, 8 January 1998.}
\title{Further evidence for the planet around 51 Pegasi}

\author{Artie P. Hatzes, William D. Cochran, \&  Eric J. Bakker}
\affil{McDonald Observatory, The University of Texas at Austin,
    Austin, TX 78712}

	{\bf The discovery$^1$ of the planet around the solar-type star
51 Pegasi  marked a watershed in the search for extrasolar planets.
Since then seven other solar-type stars have been
discovered$^{2-6}$, of which several have surprisingly short orbital periods,
like the planet around 51 Peg. 
These planets were detected using the indirect technique of measuring
variations in the Doppler shifts of lines in the spectra of the primary stars.
But it is possible that oscillations of the stars themselves (or other effects)
could mimic the signature of the planets, particularly around the short-period
planets.
The apparent lack of spectral$^7$ and brightness variations$^8$, however, led
to widespread acceptance that there is a planet around 51~Peg.
This conclusion was challenged by the observation$^9$ of systematic variations
in the line profile shapes of 51~Peg, which suggested stellar
oscillations$^{10}$.
If these observations are correct, then there is no need to invoke a planet
around 51~Peg to explain the data.
Here we report observations of 51~Peg at a much higher spectral resolution than
those in ref.~9, in which we find no evidence for systematic changes in the
line shapes.
The data are most consistent with a planetary companion to 51~Peg.
}

Observations were made using the coud{\'e} echelle
spectrograph$^{11}$ of the 2.7-m Harlan J. Smith telescope at the McDonald
Observatory.
This instrument was used in the configuration that provided
a resolving power ($\lambda/{\Delta\lambda}$) of 220,000, or more than twice
that of the data obtained by Gray$^9$,
and  with a wavelength coverage of several hundred (non-contiguous)
{\AA}ngstroms. Approximately 120 observations were made over the course of
18 nights in July -- September 1997.  On each night 4--10 observations 
of 51 Peg were made  with a typical signal-to-noise ratio of $\sim$200
for each spectrum.
Information about the spectral line shapes was
extracted using the line bisector which is the locus of the midpoints
of the stellar absorption line from the line core up to the continuum.
The velocity span  and the bisector curvature were both measured. 
The span is the velocity difference between the bisector points
taken at flux levels of 0.48 and 0.85 of the continuum  level
and the curvature is the difference 
in the velocity span  between the top half (measured between  flux points
0.64 and 0.85 of the
continuum) and bottom half (measured between 
flux points 0.42 and 0.64 of the continuum) 
of the line bisector. 
(See Gray$^{9}$  and Gray \& Hatzes$^{10}$  for illustrations
of a typical bisector as well as definitions of the
velocity span and curvature.)
Both the span and the bisector curvature were examined because it is possible
for spectral variations to have large changes in the curvature without
significant changes in the velocity span and vice-versa.
The spectral coverage of our
data included the 6,252.57-{\AA} Fe I line used by Gray as well
as 4 other spectral features of comparable
line strength that were suitable for bisector measurements:
5,296.7-{\AA} Cr I, 5,554.93-{\AA} Fe I, 5,639.95-{\AA} Fe I, and
6141.75-{\AA} Fe I. If the kinematic nonradial pulsation model
proposed by Gray \& Hatzes$^{10}$ is correct, then changes in the lines
bisectors for spectral lines of equal strength should be the same, assuming
that the spectral lines are formed at the same depth in the stellar atmosphere.
Averaging bisectors from  several line should decrease the noise level in the
measurements.

Figure 1 shows the results of our velocity span measurements.
The top panel shows the nightly average of the
velocity span of 
6,252.57-{\AA} Fe I, the same feature measured by Gray$^9$, as
a function of phase. Phases were computed using the orbital
ephemeris of Marcy et al.$^{12}$ ( a period of 4.2311 d and an
epoch of Julian day 2449797.773) which was also used by Gray and Hatzes$^{10}$.
The solid line represents the predicted sinusoidal variation of Gray$^9$.
Each error bar is the standard deviation of individual nightly
measurements weighted by the square root of the number of measurements
used in the  average.
The r.m.s. scatter of the measurements about the mean is $\sim25$\,m\,s$^{-1}$,
which is half of the amplitude of the variations found in 51~Peg by Gray$^9$
and is comparable to the r.m.s. scatter for non-variable stars.
There may be a hint of sinusoidal variability
in the measurements, but 180$^\circ$ out-of-phase with
the curve predicted by Gray which  suggests that this variability is
due to noise. This is confirmed in the lower panel which shows
the average velocity span
for all five spectral lines that were examined. The errors
shown represent the weighted standard deviation of the span
measurements from the five spectral lines. The scatter (standard deviation) of
the points shown is $\sim7$\,m\,s$^{-1}$, well below the amplitude
of the Gray prediction (line).

	Figure 2 shows the result of our bisector curvature measurements.
Once again, the top panel is for the 6,252.57-{\AA} Fe I feature
and the bottom panel is the average for all spectral lines. Errors were
computed in the same manner as for Figure 1. The curve represents
the fit to the variations from Gray \& Hatzes$^{10}$. Spectral variations,
above the noise, 
do not seem to be present in either the measurements for the individual
6,252.57-{\AA} Fe I or in the average curvature for all 5 spectral lines.
The standard deviation of the averaged measurements (lower panel) is
about 12 {\ms}.

	The best direct comparison of our
data to Gray's is made with the 6,252.57-{\AA} Fe I, the only featured
in common with both data sets. Unfortunately, these alone cannot
refute the Gray result with complete certainty. Phasing the data
to the 4.2311 day radial velocity period does show a 180$^\circ$ phase
shift between
the two measurements and tentatively this would seem to contradict
his claim of spectral variability. However, the scatter of our 
measurements are comparable to the amplitude of the variations which
may be present.
A periodogram analysis using our data combined
with Gray's shows this period as the highest peak,
and the inclusion of our data reduces the power at secondary peaks.
We note, however, that a periodogram analysis (not shown) of our data only does
not show any significant power (false alarm probability $\sim$50\%) at the
appropriate period.
A more detailed analysis and its implications will be presented elsewhere.

The averaged bisector measurements, on the other hand,
almost certainly refute the Gray result, but not without caveats.
Using the averaged measurements from different lines is valid so long
as the spectral variations are the same (and in phase) for all spectral
lines used in the analysis.
This is the case for the kinematic pulsation model
examined by Gray \& Hatzes.$^{10}$ which is only true if all spectral lines
are formed at the same depth of the stellar atmosphere and have
similar excitation potentials.   However, only the 5,296.70{\AA} Cr~I line has
a significantly different excitation petentail (0.98ev) that the other spectra
lines (3.6-4.5ev).
If 51 Pegasi is really a long period
pulsating star, then the vertical (radial) wavelength of the 
pulsations should be very short and spectral lines formed at different
depths of the stellar atmosphere may show different behavior
with phase in the spectral line shapes.
 If our line averaging technique is valid, then the spectral
variability reported by Gray is most likely an artifact of noise.
(Indeed, a Fourier analysis of the data in ref.~9 showed that there was
about a 1 in 300 chance that pure noise could produce his 
observed variability.$^{10}$)
As the spectral variability of 51~Peg is not confirmed, the nonradial
pulsation model for this star proposed by Gray \& Hatzes$^{10}$ is wrong.
We will present elsewhere a more detailed analysis of our work, including line
depth ratio measurements, variability of which was also reported by Gray$^9$.

Stellar RV measurements provide only an indirect means of finding extra-solar
planets and until there is a direct detection (either through imaging
or from the spectral signature of the planet) we can never be
absolutely certain that 51 Peg indeed has a planet. But
in the light of all observational evidence now available for 51 Peg,
including the measurements presented here, the planet hypothesis is the
simplest and most plausible explanation for the variability
reported by Mayor \& Queloz$^{1}$.

{\small Received 13 November, accepted 25 November 1997.}

\centerline{ {\bf REFERENCES}}
\begin{enumerate}
\itemsep -10pt

\item Mayor, M. \& Queloz, D. A Jupiter-mass companion
to a solar-type star. {\it Nature}, {\bf 378}, 355-359 (1995).

\item Marcy, G.W. \& Butler, R.P. A planetary companion to 70 Virginis
{\it Astrophys. J.}  {\bf 464}, L147-L151 (1996).

\item Butler, R.P. \& Marcy, G.W.  A planet orbiting 47 Ursae Majoris.
{\it Astrophys. J.}  {\bf 464}, L153-L156 (1996).

\item Butler, R.P., Marcy, G.W., Williams, E. Hauser, H. \& Shirts, P.
Three new `51 Peg-type' planets. {\it Astrophys. J.}  {\bf 474}, L115-L
(1997).

\item Cochran, W.D., Hatzes, A.P., Butler, R.P., \& Marcy, G.W.
The discovery of a planetary companion to 16 Cygni B. 
{\it Astrophys. J.}  {\bf 483}, 457-  (1997).

\item Noyes, R.W., Jha, S., Korzennik, S.G.,
Krockenberger, M., Nisenson, P., Brown, T.M., Kennelly, E.J., 
Horner, S.D. A planet orbiting the star $\rho$ Coronae Borealis.
{\it Astrophys. J.}  {\bf 483}, L111- (1997).

\item Hatzes, A.P., Cochran, W.D., \& Johns-Krull, C.M.
Testing the planet hypothesis: a search for variability in the 
spectral line shapes of 51 Peg. {\it Astrophys. J.}  {\bf 478},
374-  (1997).

\item Henry, G.W., Baliunas, S.L.
Donahue, R.A., Soon, W.H., Saar, S.H. {\it Astrophys. J.}  {\bf 474},
503- , (1997).

\item Gray, D.F. Absence of a planetary signature in the spectra
of the star 51 Pegasi. {\it Nature}, {\bf 385}, 795-796 (1997).

\item Gray, D.F. \& Hatzes, A.P. Nonradial oscillation in the
solar-temperature star 51 Pegasi. {\it Astrophys. J.}  {\bf 490},
412--424 (1997).

\item Tull, R.G., MacQueen, P.J.,
Sneden, C., \& Lambert, D.L. The high-resolution
cross-dispersed echelle white pupil spectrometer
of the McDonald Observatory 2.7-m telescope.
{\it PASP}, {\bf 107}, 251- (1995).

\item
Marcy, G.W., Butler, R.P., Williams, E., Bildsten, L., Graham,
J.R., Ghez, A.M., Jernigan, J.G. The planet
around 51 Pegasi.   {\it Astrophys. J.}  {\bf 481}, 926-  (1997).

\end{enumerate}

\begin{figure}
\plotone{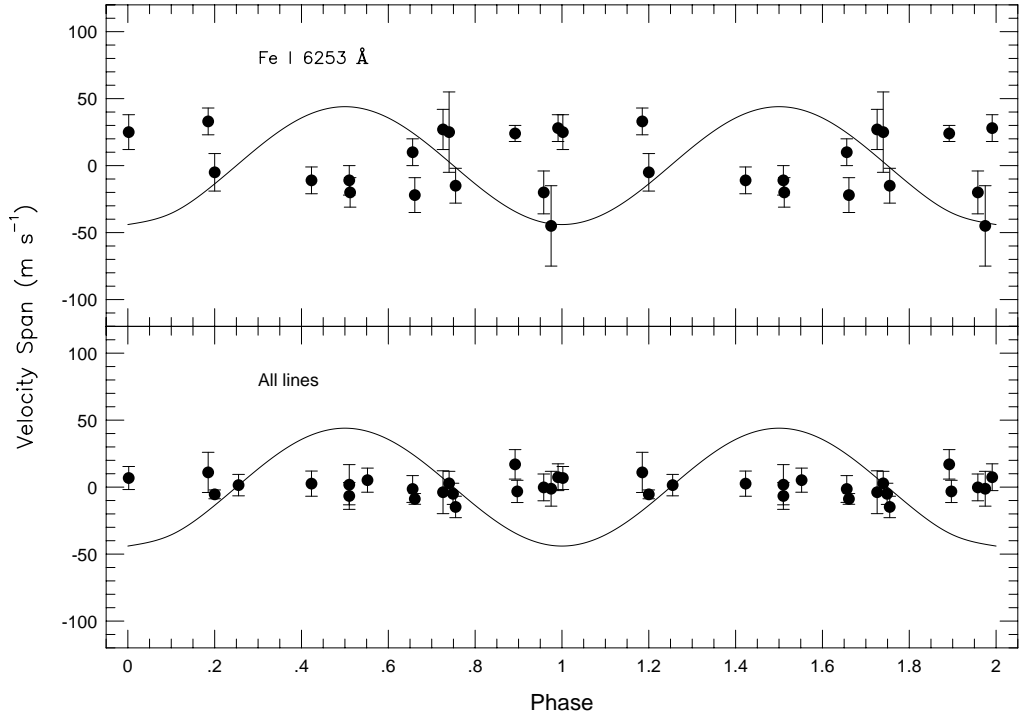}
\caption[fig1.ps] {Velocity span measurements of 51~Peg.  Top panel,
the velocity span of 6,252.57 Fe I as a function of phase (data points)
according to the ephemeris of Marcy et al.$^{12}$.
The solid line represents the variations reported by Gray$^{9}$.
Bottom panel, the average velocity span for the spectral lines:
5,296.7-{\AA} Cr I, 5,554.93-{\AA} Fe I, 5,639.95-{\AA} Fe I,
6141.75-{\AA} Fe I, and 6,252.57 Fe I. Again, the line represents
the predicted curve from  Gray$^{9}$.
\label{bisect1}}
\end{figure}

\begin{figure}
\plotone{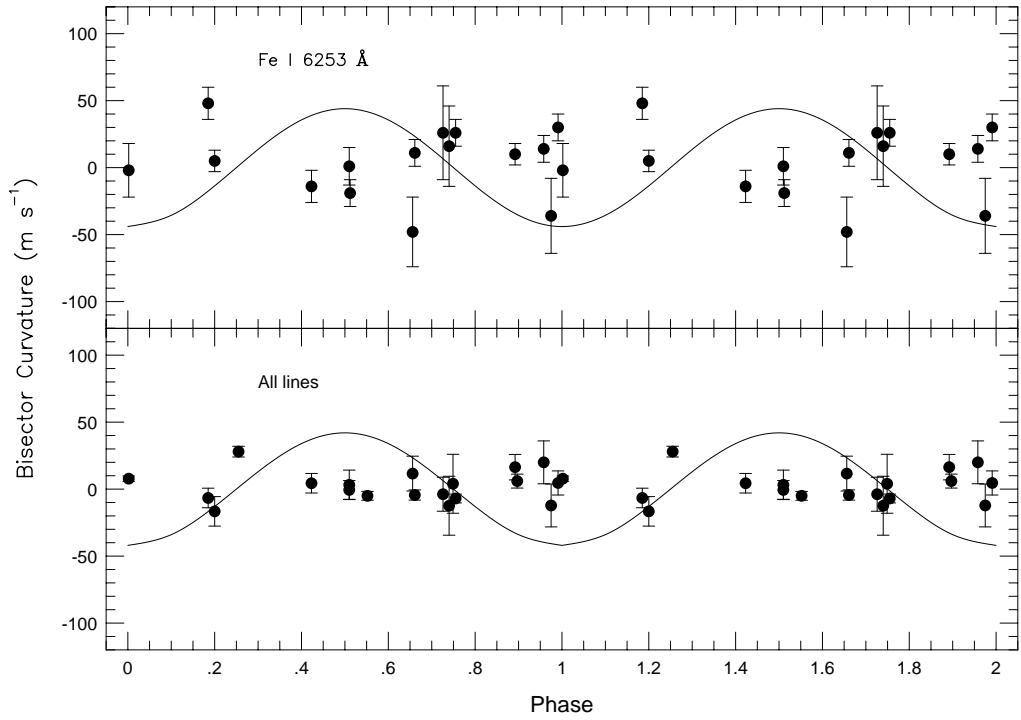}
\caption[fig2.ps] { Bisector curve measurements of 51~Peg.
Top panel, the bisector curvature of 
6,252.57 Fe I as a function of phase. The solid line represents the 
variations reported by Gray \& Hatzes$^{10}$.
Bottom panel, the average bisector curvature for five  spectral lines
along with the predicted curve of Gray \& Hatzes$^{10}$.
\label{bisect2}}
\end{figure}

\end{document}